\documentclass[12pt]{article}

\usepackage{verbatim}
\usepackage{amsmath}
\usepackage{amssymb}
\usepackage{graphicx}
\usepackage{cite}
\usepackage{subfig}
\usepackage{setspace}
\usepackage{url}
\usepackage[percent]{overpic}
\usepackage{slashed}
\usepackage{authblk}

\usepackage{xspace}

\usepackage{fullpage}
\usepackage{hyperref}
\def\tev{\,{\rm TeV}}
\def\gev{\,{\rm GeV}}

\def\to{\rightarrow}

\title{The CP-Violating pMSSM at the Intensity Frontier{\footnote {Contributed to the Community Summer Study 2013, 
Minneapolis, MN July 29 - August 6, 2013}}}

\date{}

\author{Joshua Berger$\rm ^a$}
\author{Matthew W. Cahill-Rowley$\rm ^a$}
\author{Diptimoy Ghosh$\rm ^b$}
\author{\mbox{JoAnne L. Hewett$\rm ^a$}}
\author{Ahmed Ismail$\rm ^a$}
\author{Thomas G. Rizzo$\rm ^a$}
\affil{$\rm ^a$SLAC National Accelerator Laboratory,\\ 
2575 Sand Hill Road, Menlo Park, CA 94025, USA
\footnote{mrowley, hewett, aismail, rizzo@slac.stanford.edu}}
\affil{$\rm ^b$INFN, Sezione di Roma, \\ 
Piazzale A. Moro 2, I-00185 Roma, Italy\footnote{diptimoy.ghosh@roma1.infn.it}}

\begin{document}

\rightline{\vbox{\halign{&#\hfil\cr
&SLAC-PUB-15747\cr
}}}

{\let\newpage\relax\maketitle}

\begin{abstract}
In this Snowmass whitepaper, we describe the impact of ongoing and
proposed intensity frontier experiments on the parameter space of the
Minimally Supersymmetric Standard Model (MSSM).  We extend a set of
phenomenological MSSM (pMSSM) models to include non-zero CP-violating
phases and study the sensitivity of various flavor observables in these
scenarios  Future electric dipole moment and rare meson decay
experiments can have a strong impact on the viability of these models that is
relatively independent of the detailed superpartner spectrum.  In particular,
we find that these experiments have the potential to probe models that are expected to escape
searches at the high-luminosity LHC.
\end{abstract}

\clearpage
The Large Hadron Collider (LHC) has reached a major milestone by discovering 
a Higgs boson\cite {LHCH}.  At present, the properties of this Higgs boson resemble
those predicted by the Standard Model, but the naturalness of the electroweak scale
remains unexplained.   
Supersymmetry (SUSY) in general, and its minimal version, the Minimally Supersymmetric 
Standard Model (MSSM) in particular, explains this scale and is among the best-motivated theory of physics
beyond the SM. Many of the LHC
searches have focused on its signatures, with null results so far, and the large LHC dataset is
pushing the limits on the scale of 
New Physics (NP) to roughly a TeV.  It is thus paramount that signatures of 
Supersymmetry be studied in all possible manners, including its indirect effects
at the intensity frontier.

If no assumptions are made about the SUSY breaking sector then 
the total number of unknown parameters (the so called soft SUSY breaking parameters) 
in the R-parity conserving version of the MSSM is large (105), and 
it becomes difficult to carry out a phenomenological analysis.
In order to circumvent this, two complementary approaches are
generally taken.  

The first common approach is to assume particular patterns 
for many of the parameters at some high scale.  The soft
parameters at the electroweak scale are then generated by
Renormalization Group (RG) evolution from the high scale.  
While these minimal models make the phenomenological analysis comparatively 
straightforward, they do not represent the full set of SUSY signatures and they
are now tightly constrained by a host of 
experimental observables and direct searches. In fact, two of the 
most commonly studied scenarios, the minimal super gravity (mSUGRA) and 
minimal gauge mediated SUSY breaking (mGMSB) models \cite {SUSYrefs},
are now severely constrained by the LHC
data\cite{Dighe:2013wfa}.

An alternate approach is to maintain ignorance of the physics at the high scale,
and to choose a pattern for the parameters at the weak
scale based on current experimental constraints.  A study of such models is only feasible
because many of these parameters  
are already tightly constrained by a host of low energy measurements. 
For example, both the Charge-Parity (CP) conserving and CP-violating 
observables in $K$, $B$ and $D$ decays, as well as lepton flavor violating 
decays and data on electric and magnetic dipole moments, already forbid large values 
of new CP-violating phases and sfermion mixing angles.

Taking the limit of no new sources of flavor- or
CP-violation leads to the general 19/20-parameter
pMSSM\cite{Djouadi:1998di}. The increased dimensionality of the
parameter space not only allows for a more unprejudiced study of SUSY,
but can also yield valuable information on `unusual' scenarios,
identify weaknesses in the current LHC analyses and provides the means to
combine results obtained from many independent SUSY-related
searches. To these ends, we have recently embarked on a detailed study
of the signatures for the pMSSM at the 7 and 8 TeV LHC, supplemented
by input from Dark Matter (DM) experiments as well as from precision
measurements of the Higgs properties{\cite{us}}. The pMSSM is
the most general version of the R-parity conserving MSSM when it is
subjected to a minimal set of experimentally-motivated guiding
principles: ($i$) No new sources of CP-violation, ($ii$) Minimal Flavor Violation at
the electroweak scale so that flavor violation is proportional to the CKM
mixing matrix elements, ($iii$) degenerate 1\textsuperscript{st} and
2\textsuperscript{nd} generation sfermion masses, and ($iv$)
negligible Yukawa couplings and A-terms for the first two
generations. In particular, no assumptions are made about physics at
high scales, e.g., the nature of SUSY breaking, in order to capture
electroweak scale phenomenology for which a UV-complete theory may not
yet exist. Imposing these principles ($i$)-($iv$) decreases the number
of free parameters in the MSSM at the TeV scale from 105 to 19 for the
case of a neutralino Lightest Supersymmetric Partner (LSP), or to 20 when the gravitino mass is included
as an additional parameter when it plays the role of the LSP.  We have
not assumed that the LSP relic density necessarily saturates the
WMAP/Planck value{\cite{Komatsu:2010fb}} in order to allow for the
possibility of multi-component dark matter. For example, the axions introduced
to solve the strong CP problem may make up a substantial amount of
dark matter. The 19/20 pMSSM parameters and the ranges of values employed in
our scans are listed in Table~\ref{ScanRanges}.  The parameters $M_{1,2}$,
$\mu$ and $A_{t,b,\tau}$ are all given a randomly chosen sign.  Like throwing darts,
to study the pMSSM we generate $3.7$ million model points in this
space (using SOFTSUSY{\cite{Allanach:2001kg}} and check for
consistency with SuSpect{\cite{Djouadi:2002ze}}), with each point 
corresponding to a specific set of values for these parameters. These
individual models are then subjected to a global set of collider,
flavor, precision measurement, dark matter and theoretical
constraints~\cite{Djouadi:1998di,us}.  Roughly $\sim$225k models with either type of
LSP survive this initial selection and can then be used for further
physics studies. Decay patterns of the SUSY partners and the extended
Higgs sector are calculated using privately modified versions of
SUSY-HIT~\cite{Djouadi:2006bz}, CalcHEP~\cite{calchep}, and
MadGraph~\cite{madgraph}.  Since our scan ranges include sparticle
masses up to 4 TeV, an upper limit chosen by kinematics to enable phenomenological
studies at the 14 TeV LHC, the neutralinos and charginos in either of
our model sets are typically very close to being in a pure electroweak
eigenstate as the off-diagonal elements of the corresponding mass
matrices are at most $\sim M_W$.

\begin{table}
\centering
\begin{tabular}{|c|c|} \hline\hline
$m_{\tilde L(e)_{1,2,3}}$ & $100 \gev - 4 \tev$ \\ 
$m_{\tilde Q(q)_{1,2}}$ & $400 \gev - 4 \tev$ \\ 
$m_{\tilde Q(q)_{3}}$ &  $200 \gev - 4 \tev$ \\
$|M_1|$ & $50 \gev - 4 \tev$ \\
$|M_2|$ & $100 \gev - 4 \tev$ \\
$|\mu|$ & $100 \gev - 4 \tev$ \\ 
$M_3$ & $400 \gev - 4 \tev$ \\ 
$|A_{t,b,\tau}|$ & $0 \gev - 4 \tev$ \\ 
$M_A$ & $100 \gev - 4 \tev$ \\ 
$\tan \beta$ & $1 - 60$ \\
$m_{3/2}$ & 1 eV$ - 1 \tev$ ($\tilde{G}$ LSP)\\
\hline\hline
\end{tabular}
\caption{Scan ranges for the 19 (20) parameters of the pMSSM with a neutralino (gravitino) LSP. The gravitino mass is scanned with 
a log prior. All other parameters are scanned with flat priors, though we expect this choice to have little qualitative impact on 
our results~\cite{Djouadi:1998di}.}
\label{ScanRanges}
\end{table}

While MFV arises naturally as a low energy limit of a sizable class of models, such as
gauge- or anomaly-mediated SUSY, new 
physics scenarios are generally expected to have new sources of flavor and CP-violation. 
In particular, new sources of CP-violation are well-motivated by the large cosmic 
baryon--anti-baryon asymmetry of our universe.  In this work, we aim
to go beyond the assumption of vanishing CP-violating phases in
the pMSSM.  This opens the door 
for a complementary approach to discovering SUSY.  LHC searches
are limited by kinematics, both in SUSY production and decay modes, and dark
matter searches are limited by the elastic coupling of dark matter and by
astrophysical uncertainties.  This provides a window of opportunity that may
only be probed at the intensity frontier.  In particular, cases in
which the superpartners are rather massive or nearly degenerate are well-suited
for flavor- and CP-violating searches.  By relaxing the first
assumption above, we can explore the 
sensitivity of several current and future intensity frontier experiments to the
pMSSM.  In addition to studying CP-violating quantities, we study in
greater detail several flavor-violating observable that, 
despite the MFV hypothesis, are sensitive to pMSSM models.


\begin{table}[!ht]
  \centering
  \begin{tabular}{l l l l}
    \hline
    Observable & SM Prediction & Current Exp. &
    Future Exp. \cite{Hewett:2012ns}\\ 
    \hline
    $d_e/e~(\text{cm})$ & $< 10^{-38}$ \cite{Pospelov:2005pr} & $<
    1.05 \times 10^{-27}$ \cite{Hudson:2011zz} & $< 3 \times
    10^{-31}$ \\
    $d_n/e~(\text{cm})$ & $\approx 10^{-32}$ \cite{Dar:2000tn} & $<
    2.6 \times 10^{-26}$ \cite{Baker:2006ts}
    & $< 10^{-28}$\\
    $\Delta a_\mu$ & $0$ & $(2.61 \pm 0.80) \times 10^{-9}$
    \cite{Beringer:1900zz} & $\pm
    0.15 \times 10^{-9}$\\
    $\text{Br}(K^0_L \to \pi^0 \nu \bar{\nu})$ & $2.8405 \times
    10^{-11}$ \cite{Crivellin:2012jv} & $< 2.6 \times 10^{-8}$ \cite{Ahn:2009gb} & $\pm 5\%$ \\
    $\text{Br}(K^+ \to \pi^+ \nu \bar{\nu})$ & $7.8190 \times
    10^{-11}$ \cite{Crivellin:2012jv} & $1.73^{+1.15}_{-1.05} \times
    10^{-10}$ \cite{Artamonov:2008qb} & $\pm 2\%$ \\
    $\text{Br}(B_u \to \tau \nu)$ & $1.1 \times 10^{-4}$ &
    $(0.72^{+0.27}_{-0.25} \pm 0.11) \times 10^{-4}$ \cite{Adachi:2012mm} & $\pm 5\%$ \\
    $\text{Br}(B \to X_s \gamma)$ & $3.15 \times 10^{-4}$ \cite{Misiak:2006zs} &
    $(3.40 \pm 0.21) \times 10^{-4}$ \cite{Beringer:1900zz} & $\pm 0.13 \times
    10^{-4}$ \\
    \hline
  \end{tabular}
  \caption{The complete set of observables studied in this work.  All
    observables are calculated using \texttt{SUSY\_FLAVOR v2.02}
    \cite{Crivellin:2012jv}.  The ranges for future experimental results
    assume that the SM expected values are observed and are based on the most aggressive
    experimental scenarios in \cite{Hewett:2012ns}.}
  \label{tab:sflav_observables}
\end{table}

Our analysis extrapolates 1000 models selected from the neutralino LSP pMSSM
sample described above to include CP-violating phases.  All of these selected models 
satisfy current experimental constraints on the flavor
observables described in Table \ref{tab:sflav_observables}, where these
observables are computed using \texttt{SUSY\_FLAVOR v2.02}
\cite{Crivellin:2012jv}.  Of these models, 500 were selected based on
the criterion that they are expected to be excluded at 95\%
CL by null results for a jets + MET search with $300~\text{fb}^{-1}$ of integrated luminosity at the LHC with 
$14~\text{TeV}$ c.m. energy \cite{Cahill-Rowley:2013yla}.  The remaining 500 models are
\emph{not} expected to be excluded by the same search channel with $3000~\text{fb}^{-1}$ of
integrated luminosity at the $14~\text{TeV}$ LHC.

This set of models is extrapolated beyond the
pMSSM by including all six CP-violating phases that are allowed in SUSY: $\phi_1 \equiv
\text{arg}(M_1)$, $\phi_2 \equiv \text{arg}(M_2)$, $\phi_\mu \equiv
\text{arg}(\mu)$, $\phi_t \equiv \text{arg}(A_t)$,
$\phi_b \equiv \text{arg}(A_b)$, and $\phi_\tau \equiv
\text{arg}(A_\tau)$.  One phase in the gaugino-Higgsino sector is
unphysical and we choose this to be the phase of $M_3$, which we set to
zero by field redefinition.  For each of the 1000 models, we generate random
values for each of these phases employing a log uniform distribution
between $10^{-6}\times \pi/2$ and $\pi/2$.  A random sign for each phase is also
selected.  1000 sets of 6 phases are generated for each model, leading
to a total sample of $10^6$ models.  For each of these models, the
observables in Table \ref{tab:sflav_observables} are re-calculated, again
using \texttt{SUSY\_FLAVOR v2.02}.  We note that signatures of these 1000 models
at energy and cosmic frontier experiments\cite{Cahill-Rowley:2013yla,Cahill-Rowley:2013dpa}
have also been studied for Snowmass, in order to facilitate comparisons across
the frontiers.

\begin{figure}[!ht]
  \centering
  \begin{tabular}{c c}
    \includegraphics[width=3.1in]{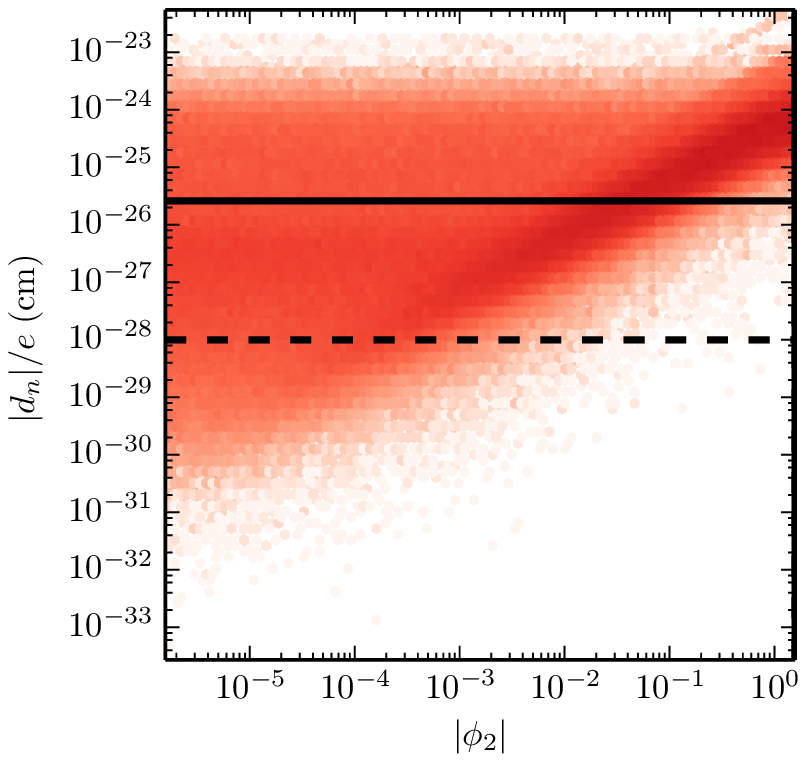} & \includegraphics[width=3.1in]{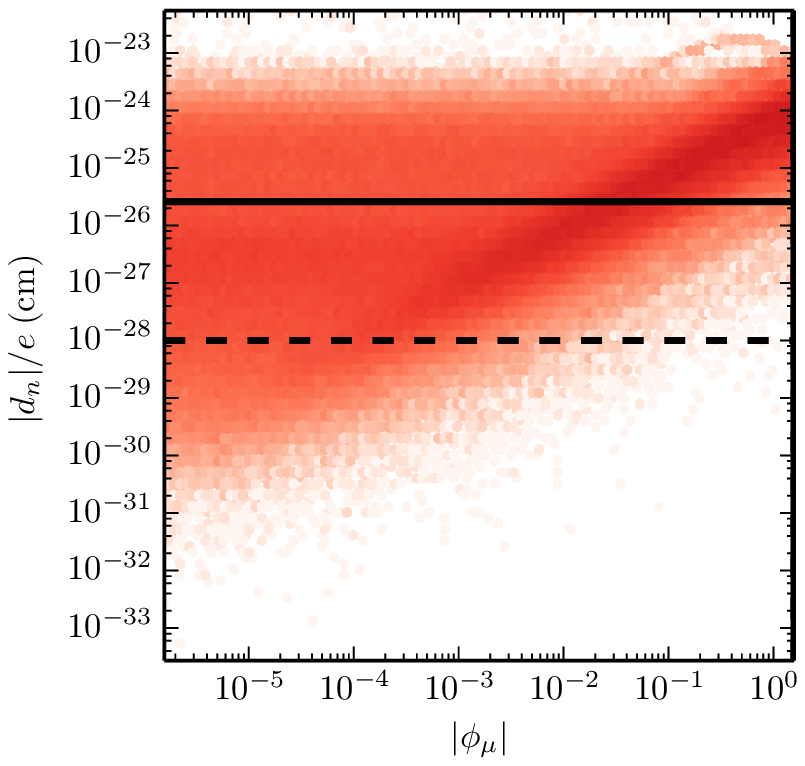} \\
    \includegraphics[width=3.1in]{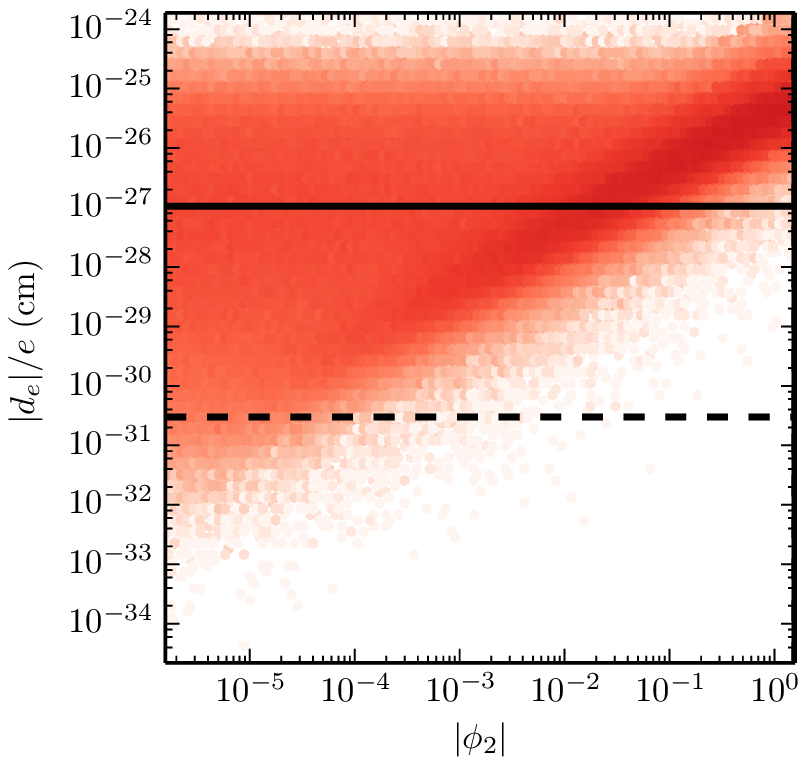} & \includegraphics[width=3.1in]{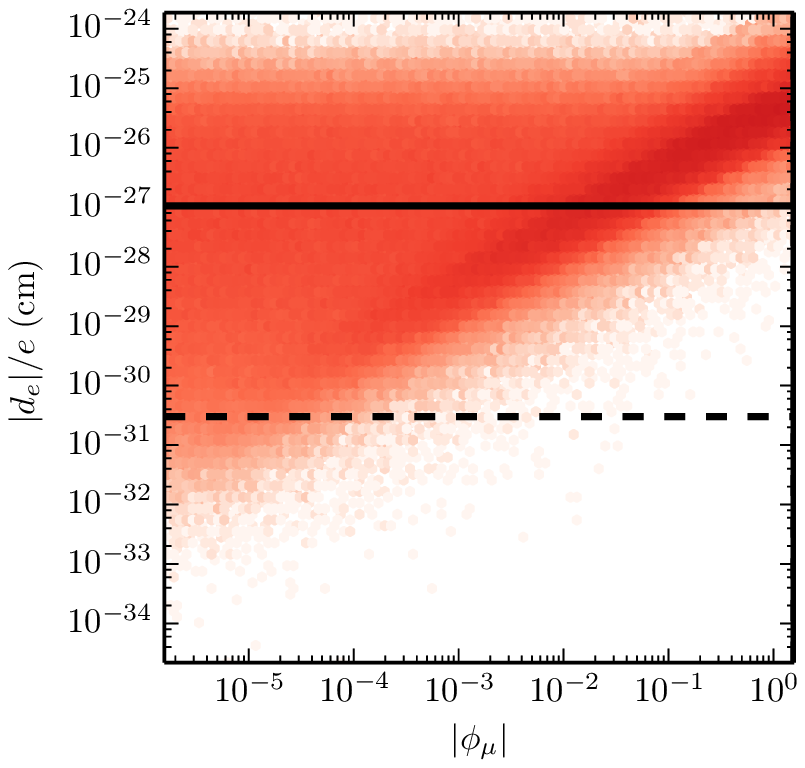}
  \end{tabular}  
  \includegraphics[width=3.1in]{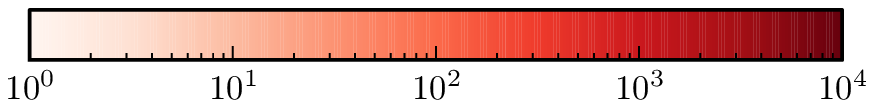}
  \caption{Results for the most sensitive EDM observables over a range
    of phases.  Solid (dashed) lines indicate current (expected future)
    experimental 2$\sigma$ bounds.}
  \label{fig:vphases}
\end{figure}
\begin{figure}[!ht]
  \centering
  \begin{tabular}{c c}
    \includegraphics[width=3.1in]{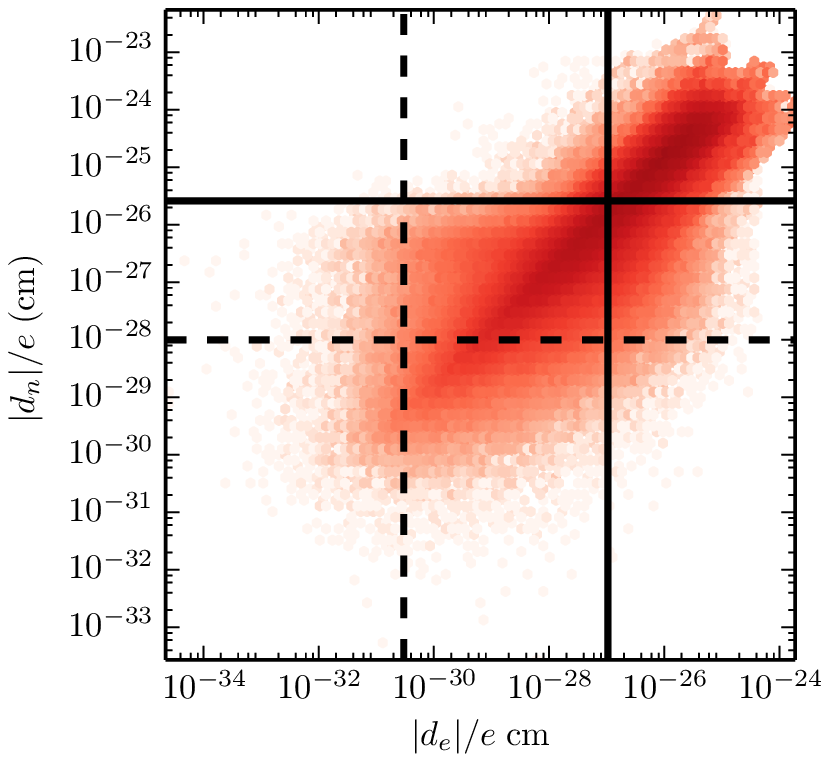} & \includegraphics[width=3.1in]{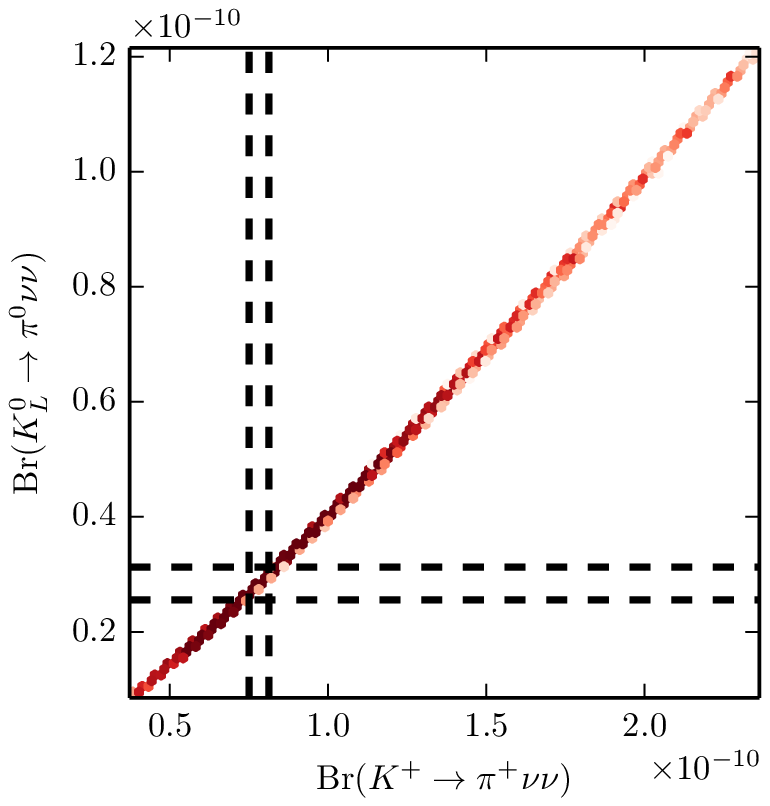} \\
    \includegraphics[width=3.1in]{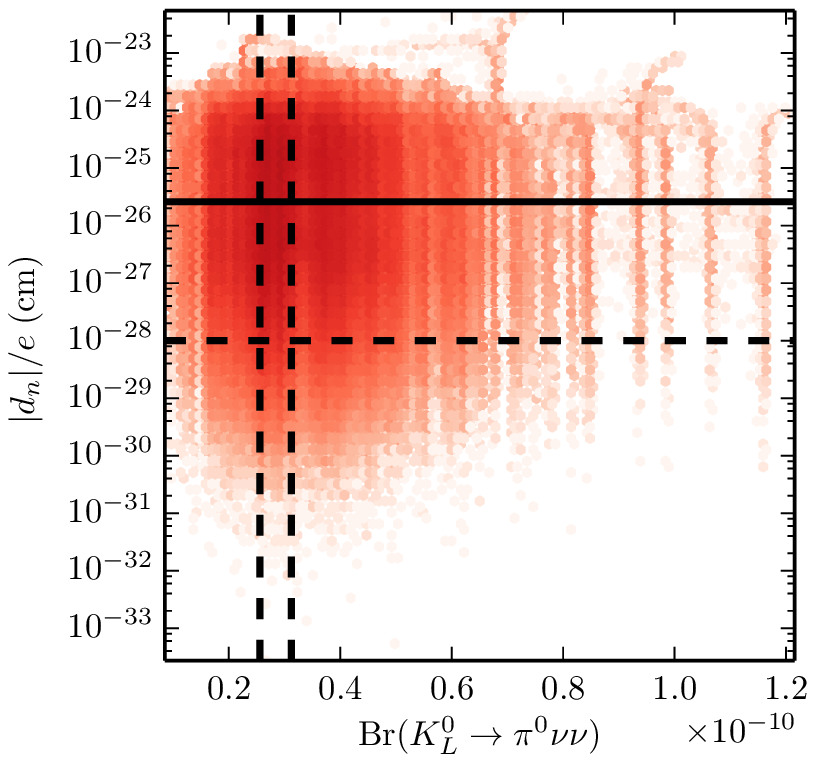} & \includegraphics[width=3.1in]{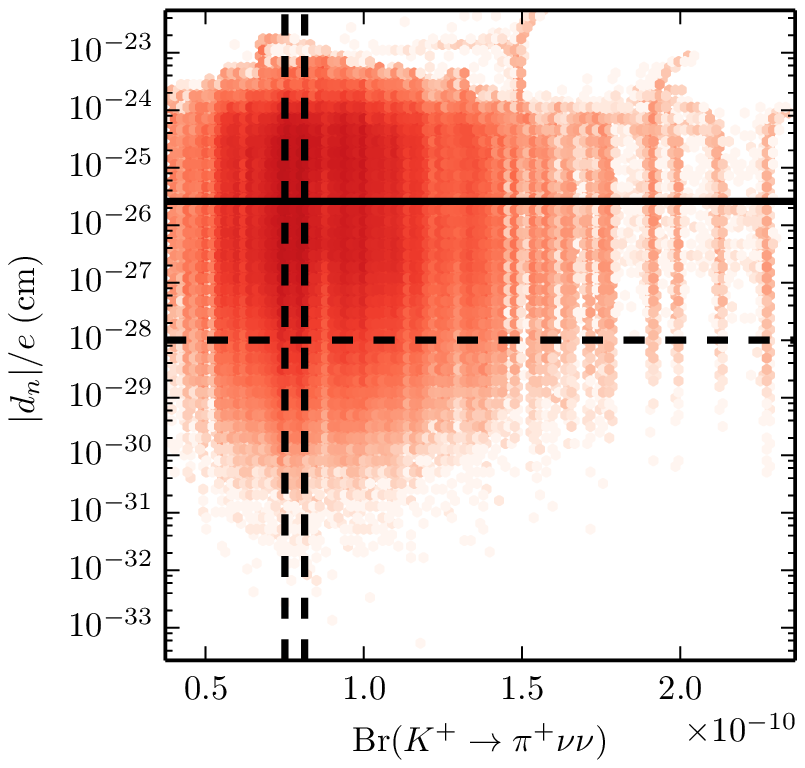}
  \end{tabular}
  \includegraphics[width=3.1in]{color_bar.eps}
  \caption{Results for the most sensitive EDM observables over a range
    of phases.  Solid (dashed) lines indicate current (expected future)
    experimental 2$\sigma$ bounds.}
  \label{fig:vobs}
\end{figure}

The results of this scan are
summarized in Figures \ref{fig:vphases} and \ref{fig:vobs} for various flavor-
and CP-violating observables.  Current
experimental bounds at the $2\sigma$ level are represented by solid
lines.  Projected future sensitivities are indicated by dashed lines,
using $2\sigma$ limits for the most aggressive experimental scenarios described in
\cite{Hewett:2012ns}.  
Both sets of limits are summarized in Table
\ref{tab:sflav_observables}.  As expected for
models with wino or Higgsino LSPs, we find that the CP-violating observables are
most sensitive to the phases of $M_2$ and $\mu$ and therefore  we only show
the dependence on these phases.
The EDM searches for the neutron and electron are seen to be complementary: there are many
models for which only one of the two most sensitive observables is large.
The branching fractions for the rare Kaon decays demonstrate the well-known MFV linear
relationship \cite{Buras:2001af}.  Any deviation from this would be a signature of
non-minimal flavor violation.  In addition, we see that there is no correlation between
the EDM values and the rare Kaon decays.  

Only the most sensitve observables are shown in these figures.
Additional weaker constraints can be obtained from ${\rm Br}(B \to X_s \gamma)$
and ${\rm Br}(B_u \to \tau \nu)$.  The distributions for the two
different sets of models, those to which the LHC is expected to have
sensitivity and those to which it is not, are comparable and we do not
separate the two sets in Figures \ref{fig:vphases} and \ref{fig:vobs}.
In all cases, we find that the expected reach
for these observables has sensitivity to models that cannot be probed at the high luminosity LHC,
provided that the CP-violating phases do not vanish.

The future for both CP-violating and flavor-violating observables is
exciting.  Experiments are slated to improve by several orders of
magnitude in sensitivity and will have a significant impact on the available
parameter space, even for models to which the LHC is unlikely to be
sensitive.  This work demonstrates the powerful and complementary role
that such probes can play in the hunt for new physics.

\clearpage
\section*{Acknowledgements}

DG is supported by the European Research Council under the European Union's 
Seventh Framework Programme (FP/2007-2013) / ERC Grant Agreement n.279972. 
DG would also like to gratefully acknowledge the hospitality and support by 
the SLAC Theory Group where part of this work was done.  This work was
supported by the Department of Energy, Contract DE-AC02-76SF00515.


\end{document}